\documentclass[ prb,twocolumn,showpacs]{revtex4}
\usepackage{amssymb}
\pdfoutput=1
\usepackage{graphicx,amsmath}
\usepackage{epstopdf}
\newcommand{\kk}{\textbf{k}}
\newcommand{\iomega}{i\omega_n}
\newcommand{\dexp}{\Delta_{exp}}
\newcommand{\dth}{\Delta}

\newcommand{\ekp}{E_{\kk +}}

\newcommand{\ro}{\rho_0}

\newcommand{\appp}{\alpha'}
\newcommand{\efp}{\epsilon_f - \appp\mu}

\newcommand{\nm}{\,nm^{-2}}

\newcommand{\ie}{\textit{i.e.} }
\newcommand{\bea}{\begin{eqnarray}}
\newcommand{\eea}{\end{eqnarray}}
\newcommand{\beq}{\begin{equation}}
\newcommand{\eeq}{\end{equation}}
\newcommand{\benu}{\begin{enumerate}}
\newcommand{\enu}{\end{enumerate}}

\newcommand{\bk}{{\bf k}}

\begin{document}

\title{Activation gap in the specific heat measurements for $^3$He bi-layers}
\date{\today}
\author{A. Ran\c con-Schweiger$^1$, A. Benlagra$^2$ and C. P\'epin$^{3,4}$}
\affiliation{$^1$LPTMC, CNRS UMR 7600, Universit\'e Pierre et Marie Curie, 75252 Paris, France.\\
$^2$Institut f\"ur Theoretische Physik, Universit\"at zu K\"oln, Z\"ulpicher Str. 77, 50937 K\"oln, Germany.\\
$^3$Institut de Physique Th\'eorique, CEA, IPhT,CNRS, URA 2306, F-91191 Gif-sur-Yvette, France.\\
$^4$International Institute of Physics, 
Universidade Federal do Rio Grande do Norte, 59078-400 Natal-RN, Brazil \\
}

\begin{abstract}
Recently, attention has been given to a system of $^3$He bi-layers where a quantum criticality similar to the one in heavy fermion compounds has been observed [Science \textbf{317}, 1356 92007)] In our previous analysis [Phys. Rev. B \textbf{79}, 045112 (2007)], based on the Kondo breakdown scenario, we addressed successfully most of the features observed in that experiment. Here, we consider the activation energy $\Delta$ observed experimentally in the specific heat measurements at low temperatures in the heavy Fermi liquid phase. Within our previous study of this system, this is identified with the gap opening when the upper hybridized band is emptied due to a strong hybridization between the nearly localized first layer and the fluid second one. We discuss the successes and limitations of our approach. An additional prediction is proposed. 
\end{abstract}
\maketitle

\section{Introduction}
In recent years, the peculiar and non standard properties observed in heavy fermion compounds (HFC)
 \cite{stewart, loneysen} have been the subject of an intense debate. In particular, the nature
of the  quantum critical point (QCP), thought to be the origin of their remarkable properties in
 a large region of the phase diagram, remains a puzzling and open problem. Many scenarios have
 been put forward to account for these properties \cite{review-piers, HMM, qsi,senthil, us, cath}.
 However, the difficulty with HFC is their complex chemical structure and the variety of physical
 phases, often competing, present in their phase diagram. These, despite a growing body of 
experimental facts, often prevent to rule in favor of a theoretical model or another.

Lately, a quantum criticality similar to the one in HFC has been observed in a much simpler 
system : $^3$He bi-layers adsorbed on a composite substrate of graphite pre-plated with two
 solidified layers of $^4$He [\onlinecite{saunders}]. Upon increasing the total coverage 
$N$ of $^3$He atoms, the system seems to undergo a quantum transition, at a critical coverage
 $N_{crit}=9.95 nm^{-2}$, separating a regime where the two layers are hybridized and form a
 bi-fluid with heavy Fermi liquid (FL) properties to a regime where they are completely 
decoupled, the first one (L1) forming a frustrated 2D magnet and the second one (L2) remaining
 fluid. At the putative QCP, there is a breakdown of the FL showing as an apparent divergence
 of the effective mass $m^*$ and a vanishing of an energy scale $T_0$ indicative of an 
effective coupling between the two layers. This is reminiscent of what happens with HFC 
except that, in this particular case, there is no long range magnetic order due to 
frustration and the energy scales are different.

This similarity has led two of the authors \cite{adel} to use one of the models proposed
 to describe quantum criticality in HFC, namely the so-called Kondo breakdown (KB) 
scenario \cite{us, cath}, to account for the remarkable features of the experiment. 
In that study, the $^3$He bi-layers system has been mapped onto an extended version of the
 periodic Anderson lattice model where the nearly localized fermions of L1 are identified 
with the $f-$ electrons and the ones of L2 with the itinerant conduction electrons. 
Remarkably, our study identifies the real QCP as occurring at a coverage of $N_I \approx 9.2 mn^{-2}$
 lower than the experimental one, whereas the experimental QCP is identified as the extrapolation to lower temperatures
 of a high energy regime of fluctuations.  The coverage $N_I  \approx 9.2 nm^{-2}$ corresponds to  the 
 abrupt increase of the magnetic susceptibility.
The QCP  in our theory  corresponds to an 
orbital-selective Mott transition at which an effective hybridization between spinons
 in L1 and conduction electrons in L2 vanishes, resulting in a solidification of L1. It has been also called Kondo Breakdown QCP. 
This approach was successful enough, 
using a small set of three  fitting parameters, to account for most of the properties observed experimentally. In particular, the apparent presence of two QCP's and the apparent absence of a critical regime in temperature have been explained. The coverage dependence of both the effective mass and the energy scale $T_0$ have been fitted in an intermediate regime of temperature. We emphasize that the mechanism invoked in our study is different from the one describing the increase of $m^*$ in the second $^3$He monolayer on bare graphite or pure 2D $^3$He [\onlinecite{boronat}]. The latter indeed does not rely on the interlayer coupling as a crucial aspect of the physics leading to the enhancement of the effective mass.

\begin{figure}[Ht]
\centering
\includegraphics[scale=0.7]{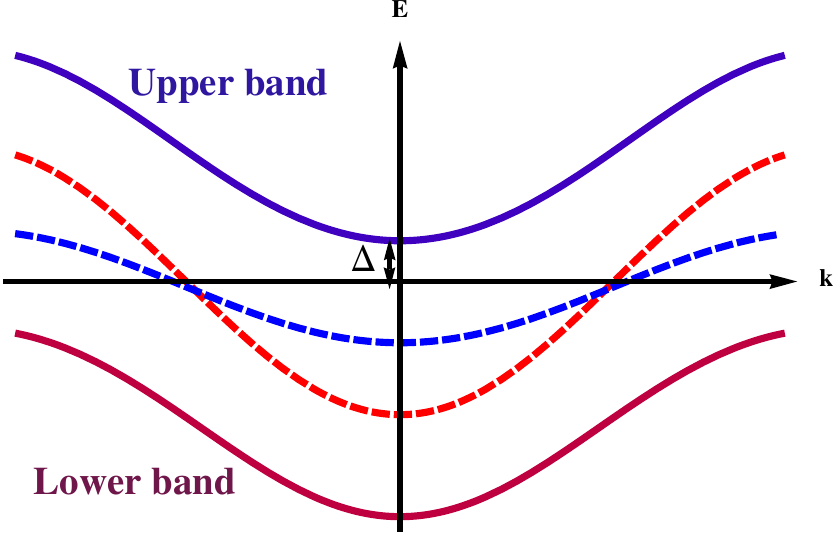} 
\centering
\caption{Sketch of the different dispersions for the hybridized bands (solid lines), the conduction electrons (upper dashed line) and the spinons (lower dashed line). The gap $\dth$ is defined as the difference between the Fermi level and the bottom of the upper band.}\label{band}
\end{figure}

\begin{figure}[h!]
\centering
\includegraphics[scale=0.7]{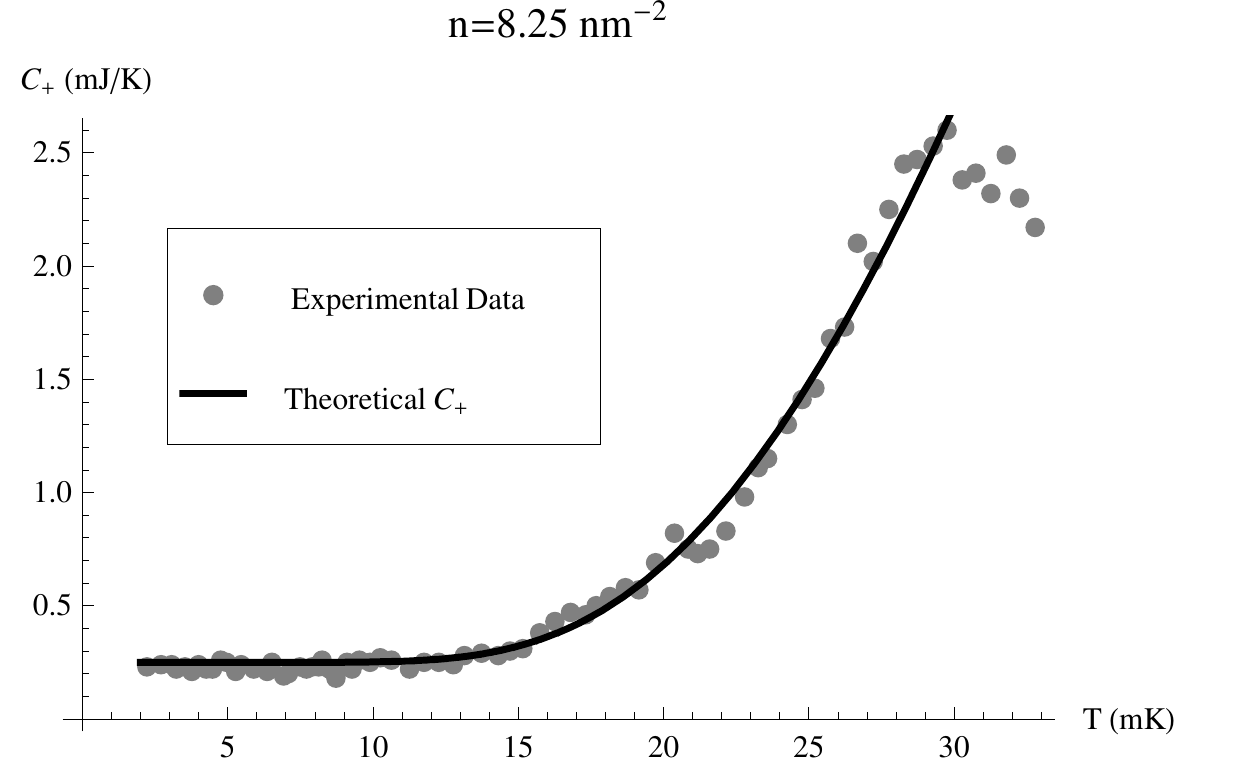} 
\centering
\caption{Comparison between $C_+$ and the  gapped part of the experimental data \cite{saunders} for $n=8.25\nm$. }\label{825}
\end{figure}

\begin{figure}[!ht]
\centering
\includegraphics[scale=0.7]{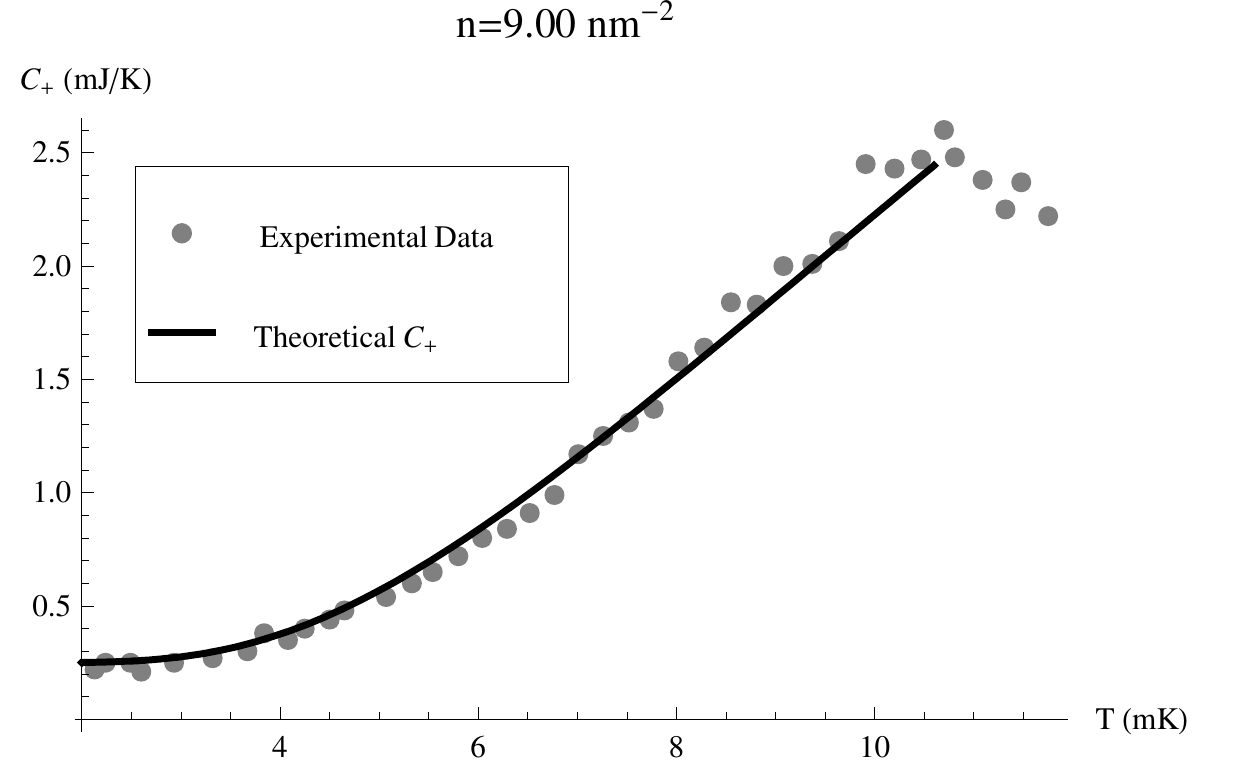} 
\centering
\caption{Comparison between $C_+$ and the  gapped part of the experimental data \cite{saunders} for $n=8.25\nm$. }\label{900}
\end{figure}

 In this paper , we consider two important additions to our previous theory. First we evaluate the specific heat contribution of the gapped band and 
compare it to the experimental data.  This allows us, in particular,  to give  a very simple interpretation for 
the activation gap observed experimentally \cite{saunders}. The presence of an excitation gap is needed to fit the specific heat measurements in the hybridized phase according to
$
C_{exp}= \gamma T + \gamma_2 e^{\frac{-\dexp}{T}},
\label{exp}
$
where the first term is usual for a FL with $\gamma$ proportional to the effective mass of the quasi-particles. The second term defines the experimental activation gap $\Delta_{exp}$. Within our theory, the gap   originates from an 
excitonic mechanism between the heavy and the light band of the system. The gap $\Delta$ is shown to open (see Fig. \ref{band}) when the upper hybridized band is emptied due to a strong hybridization between the nearly localized L1 and the fluid L2.
Our analysis reproduces successfully the coverage dependence of the  activation gap, with a good quality fit to the experimental data ( see Figures \ref{825} and \ref{900} ).
The amplitude of the gap, however, departs from the   observed experimental one by a factor of ten. Some discrepancy in the amplitude is to be expected, considering that the theory used an approximate Eliashberg scheme. 
Another analysis 
of this experiment has been carried out in ref. [\onlinecite{assaad}]. It is based on a cluster 
dynamical mean field approach and leads to the identification of the transition observed 
experimentally with a band-selective Mott transition at which the upper band is gapped beyond 
the chemical potential. It accounts for the qualitative behavior of the effective mass and the 
energy scale $T_{0}$.  However, the excitation gap as well as  the absence of a critical regime in temperature are not 
discussed in this study.

As a second addition to our previous study, we make a simple  {\it quantitative } prediction concerning the Weiss term $\theta$ in the magnetization behavior observed using NMR above $T_0$\cite{saunders}. The latter writes
\beq
M(T>T_0)=\frac{C}{T-\theta}.
\eeq
NMR providing a local measure of the nuclear magnetic susceptibility, we attribute the Weiss term to the Kondo coupling between
 the two layers. Within our model, and by definition of the Kondo Breakdown QCP, these two layers decouple at the {\it real}  QCP, at a coverage
  a bit smaller than the QCP induced by the extrapolation of the quantum fluctuation regime. Accordingly, 
the Weiss term $\theta$ should vanish  precisely at the coverage
 \begin{equation}
N_I \approx 9.2 nm^{-2}\end{equation}%
 where the real  QCP is theoretically situated . 
This quantitative prediction is simple enough to be checked in future experiments on this system.


We start with the evaluation of the specific heat contribution.
The mean-field theory considered in our study\cite{adel} consists of a two bands model with an effective hybridization $V_{eff}\equiv V b$, where $V$ is the bare hybridization and $b$ is the expectation value for the holon operator, between the spinons and the conduction electrons. Diagonalization of the corresponding Hamiltonian results in two upper (+) and lower (-) hybridized bands with dispersions $E_{\bk \pm}$. 

As discussed in our previous study of this system \cite{adel}, the bare hybridization at the QCP is already very strong compared to other energies of the model. Further, at the QCP, the $f$ band is half-filled and the upper band is constrained to stay below the Fermi level. However, as soon as the effective hybridization sets in strongly, we enter very soon a regime where  that band is empty and a gap opens. This happens in the hybridized phase before the QCP is reached, which is consistent with the experimental observation\cite{saunders}. The gap is defined simply as
\bea
\Delta & \equiv & E_+(-D) \nonumber \\
& = & \frac{1}{2} \left [ -(1+\alpha^\prime)D + \epsilon_f -\mu  \right . \nonumber \\ 
& & \left .  + \sqrt{(-(1-\alpha^\prime)D -\epsilon_f-\mu)^2+4 V_{eff}^2}\right ].
\label{gap}
\eea
Here $D$ is the half bandwidth of the conduction electrons, $\epsilon_f$ and $\mu$ are the chemical potentials of the $f-$spinons and $c-$electrons respectively and $\alpha^\prime$ is the effective ratio between the bandwidths of the two fermionic species.

The gapped band contribution to the mean-field free energy reads 
\bea
F_+&=&-2T\sum_{\kk,\iomega} \ln(-\iomega +\ekp), \label{F_expression}
\eea
where $\beta = 1/T$.

\begin{figure}[h!]
\centering
\includegraphics[scale=0.7]{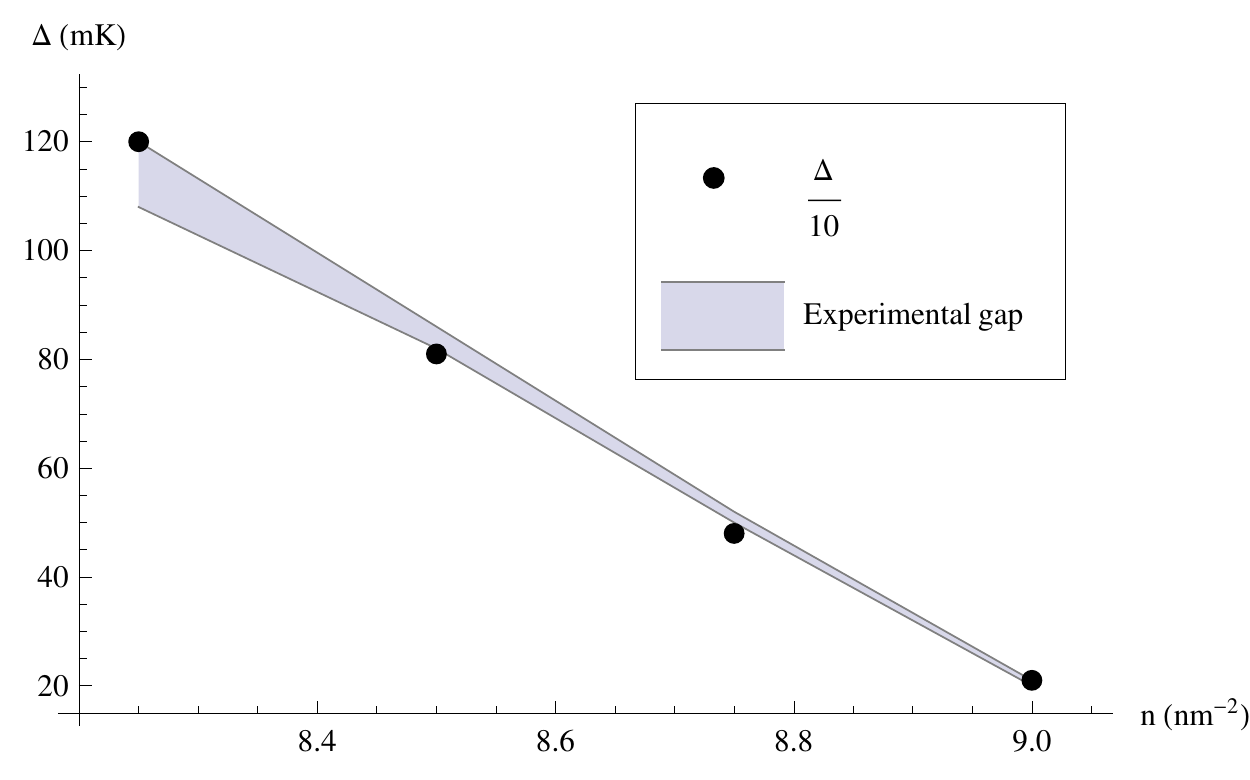} 
\centering
\caption{Comparison between $\dth$ and the experimental gap $\dexp$ \cite{saunders} for different doping.}\label{compdth}
\label{fig:gap}
\end{figure}

The evaluation of  (\ref{F_expression}) is straightforward. It is easily shown that at very low temperatures, such that $\Delta \gg T$, the specific heat contribution of the gapped band simplifies to 
\begin{equation}
C_+= \frac{\lambda}{T}e^{\frac{-\dth}{ T}},
\label{C_final}
\end{equation}
where
$$\lambda \equiv \dfrac{\ro}{\appp}\dth^2[1+\appp + \dfrac{1-\appp}{\sqrt{4\appp V_{eff}^2}}(\efp - (1-\appp ) \dth)],$$
$\rho_0$ being the density of states at the Fermi level of the conduction electrons. The expression (\ref{C_final}) is our final result. Notice that the functional form there is different from the one used for $C_{expt}$ by the experimentalists. In particular, the coefficient of the exponential depends on temperature as $1/T$.
In the following, we discuss  our results.
The theoretical gap is given by the expression (\ref{gap})
It turns out that it has the same coverage dependence, up to a constant factor of ten, of the
one extracted experimentally from $C_{expt}$ \cite{saunders} (see Fig. \ref{fig:gap})

As emphasized in the introduction, the values of the experimental gap depend strongly on the 
fitting function used by the experimentalists to capture the low temperature behavior of the 
specific heat. Thus, what has to be done is to try fitting the gapped part of the specific 
heat using the expression (\ref{C_final}) obtained within our model. 
We consider now the gapped part of the experimental data for specific heat, $
C_g(T) = C_{exp}(T) -  \gamma T,
$
we find its temperature dependence rather well reproduced by our analytical form (\ref{C_final}),
 provided the values of the gap $\Delta$ are corrected by a constant factor of ten. Figures \ref{825}-\ref{900}
 below, show this perfect agreement for 
coverages $N=8.25, 9.00 \,\,\, nm^{-2}$. 

\begin{figure}[!ht]
\centering
\includegraphics[scale=0.7]{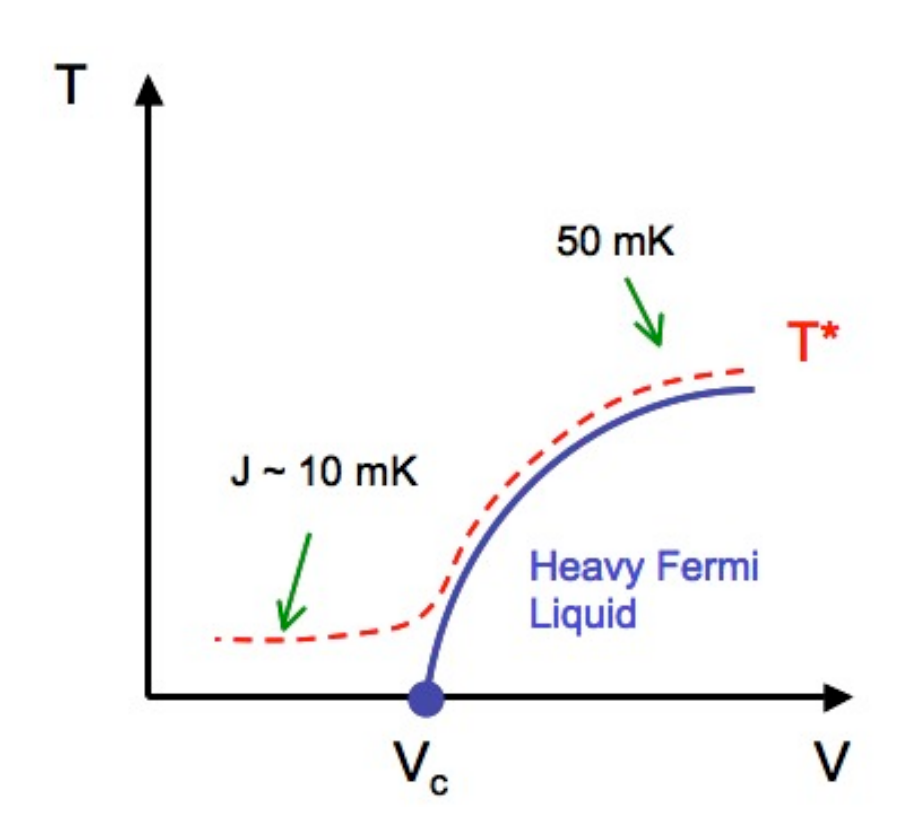} 
\centering
\caption{  Phase diagram of the  Kondo Breakdown QCP. There are two successive quenching of the entropy, one at $T_0$ due to the formation of the spin liquid and the phase transition temperature. The Kondo scale is associated to the  phase transition to the heavy Fermi liquid and should vanish at  the QCP.}\label{KB}
\end{figure}

The analytic form (\ref{C_final}), deduced form our model, seems indeed to be adequate to describe the observed low temperature behavior
of the gapped specific heat. One can refine the effective magnitude of the gap $\Delta$ assumimg that it
 has nodes in the momentum space, \ie it doesn't open uniformly along 
the Fermi surface. Indeed, the layer L1 solidifies into a commensurate lattice with respect to 
the underlying substrate which has a triangular lattice\cite{saunders}. Thus, the bare 
hybridization $V$ could be momentum dependent, following one irreducible representation of the 
lattice symmetry group, whereas it is assumed to be local in our model \cite{adel}. 
Investigation in this direction and its possible implication have not been performed yet.
In this paper as well as in a series of  previous studies \cite{adel} we have used the KB  theory of quantum criticality in order to study 
 the phase transition in  $^3$He bi-layers.  One important feature of our model,
 generally poorly understood, is the key role of frustration around the phase 
transition. In the Kondo lattice model, there are two main  mechanisms to quench entropy 
(see Fig. \ref{KB}).  One is through the formation of Kondo singlets, which after merging 
together are forming the heavy Fermi liquid, and the  other one is through entanglement of
 the spins between themselves, resulting  in the formation of the spin liquid\cite{note1}. 
 The quenching of the entropy through spin liquid formation is highly sensitive to the presence
 of frustration in the system, which is  indeed present in $^3$He bi-layers due to ring exchange interactions\cite{ roger,misguich}.
\vspace{-0.00 cm}
 We want to emphasize that in  our KB model for  $^3$He bi-layers,  the QCP is driven
 by the spin liquid and not by the formation of the Kondo singlets. The  entropy is 
quenched through the formation of a spin liquid  while the phase transition towards the heavy 
Fermi liquid is associated with the breakdown of the Kondo effect.
 
This observation enables us to make a  quantitative theoretical prediction. 
 One might be able to extract the
  Kondo scale experimentally, for example through a 
Nuclear Magnetic Resonance experiment where the relaxation of the first layer to the bath  
of free fermions made of the second layer is predominant.  A Curie-Weiss law for the susceptibility
 is expected there where the Curie-Weiss term $\theta $ can be interpreted as the Kondo scale of 
the system.
    
One of the results of our previous investigation of this model \cite{adel} is that two apparent QCPs are present.  We identified the experimental one with the extrapolation of a finite temperature regime of fluctuations. We claim however that the real QCP is occurring ``before'' the experimental one, when coverage is increased.
  
   If our model is correct, the scale $\theta$ should go to zero at the ``real'' QCP,  and not at the experimental one.
   Concretely it means that  within our theory, $\theta$  vanishes for this experimental set up at the coverage 
   \[ n_I \approx 9.2 nm ^{-2}.\]
   It is a strong prediction of our model and we hope that it can be tested experimentally in a very near future.
The system studied, $^3$He bi-layers, in our serie of papers, ref. [\onlinecite{adel}] and the present one, is one of the simplest physical ones, with negligible spin-orbit interactiuon and no crystal-field interactions. The beauty of the experiment in ref. [\onlinecite{saunders}] is that it shows that such a system has qualitatively the same physics as HFC and may serve thus as a test physical system for theories of quantum criticality in these complex systems.

In this manuscript, we addressed one final feature of the experiment on this system, namely
 the presence of an activation gap in the low temperatures behavior of the specific heat in 
the HFL phase. We derived an analytic expression for the specific heat within our model, different
 from the one used by experimentalists. The behavior of the specific heat at low temperature is 
sucessfully reproduced
as well as the coverage dependence of the gap, albeit the latter has to be corrected by a 
constant factor of ten. 
 This is expected within the Eliashberg treatment used here and may be a hint for a refinement of our model or a re-evaluation of the parameters used
 in our previous study.

The KB model is successful and strong enough to account for main of the features of the experiment
 and make some quantitative predictions to be tested in future experiments on this system. 
It is now up to new experimental data to validate it or not and to different theoretical models to propose
an alternative interpretation of the experiment. 

\begin{acknowledgments}
This work was supported by the French National Grant ANR26ECCEZZZ. We acknowledge fruitful discussions with J. Saunders and I. Paul.
\end{acknowledgments}


\begin{thebibliography}{99}
\bibitem{note1}  At lower temperature of course, the occurrence of phase transitions, of magnetic type for exemple, can quench the entropy further.
\end{thebibliography}
\end{document}